# Phase Transition of Dynamical Herd Behaviors in Financial Markets

Kyungsik Kim* and Seong-Min Yoon [a]

*Department of Physics, Pukyong National University, Pusan 608-737, Korea*

[a]*Division of Economics, Pukyong National University, Pusan 608-737, Korea*

ABSTRACT

We study the phase transition of dynamical herd behaviors for the yen-dollar exchange rate in the Japanese financial market. It is obtained that the probability distribution of returns satisfies the power-law behavior $p(R) \sim R^{-\beta}$ with three different values of the scaling exponent $\beta$ = 3.11 (one time lag $\tau$ = 1 minute), 2.81 (30 minutes), and 2.29 (1 hour). The crash regime in which the probabilty density increases with the increasing return appears in the case of $\tau < 30$ minutes, while it occurs no financial crash at $\tau > 30$ minutes. it is especially obtained that our dynamical herd behavior exhibits the phase transition at one time lag $\tau$ = 30 minutes.

*Corresponding author. Tel: +82-51-620-6354; Fax: +82-51-611-6357.
E-mail address: kskim@pknu.ac.kr.

The microscopic models in financial markets have recently received a lot of attention [1-4]. Of many crucial models, the main concentration is on the herding multiagent model [5,6] and the related percolation models [7,8], the democracy and dictatorship model [9], the crowd-anticrowd theory, the self-organized dynamical model [10], the cut and paste model, the fragmentation and coagulation model [11]. One of microscopic models in the self-organized phenomena is the herding model [12,13], in which some degrees of coordination among a group of agents share the same information or the same rumor and make a common decision in order to create and produce the returns. There are three important reasons to be influenced into the herd behavior: First, it exists the crash model that the herds may be occurred by the biased information between investors. Second, the return structure of fund managers may be sensitive to the herd behavior, since bank and stock company influence powerfully to investors. Lastly, fund manager and market analyst may play a crucial role to essentially determine the investment, in order to maintain their reputation and credit. It is particularly of interest for the herd model to search for the bubbles and crashes in econophysical system. The probability distribution of returns shows a power-law behavior for the herding parameter below a critical value, but the financial crashes yields an increase in which the probability of large returns exists for the herding parameter larger than the critical value.

The theoretical and numerical analyses for the volume of bond futures transacted at Korean futures exchange market have presented in the previous work [14]. This treated mainly with the number of transactions for two different delivery dates and found the decay functions for survival probabilities [15] in the Korean bond futures. We also argued the tick dynamical behavior of the Korean bond futures price using the range over standard deviation or the R/S analysis in the futures exchange market [16]. Skjeltorp [17] has shown that there exists the persistence caused by long-memory in the time series on Norwegian and US stock markets. The numerical analyses based on multifractal Hurst exponent and the price-price correlation function have used for the long-run memory effects. The crash regime for the yen-dollar exchange rate appears, when the herding parameter $h$ satisfies $h \geq 2.33$ in the case of one time lag $\tau = 1$ hour [18]. Our purpose of this paper is to investigate the herd behavior and the phase transition for the yen-dollar exchange rate in the Japanese financial market. Our obtained result will compared with other numerical calculation.

First of all, we analyze tick data of the yen-dollar exchange rate for the period 2nd January 2001 - 31th December 2001. We show minutely time series of the price return for the yen-dollar exchange rate in Fig. 1. The price return $R(t)$ can be defined as

$$R(t) = \ln \frac{P(t+\tau)}{P(t)} \qquad (1)$$

where $P(t)$ is the price of the yen-dollar exchange rate at time $t$ and $\tau$ denotes one time lag between transacted times.

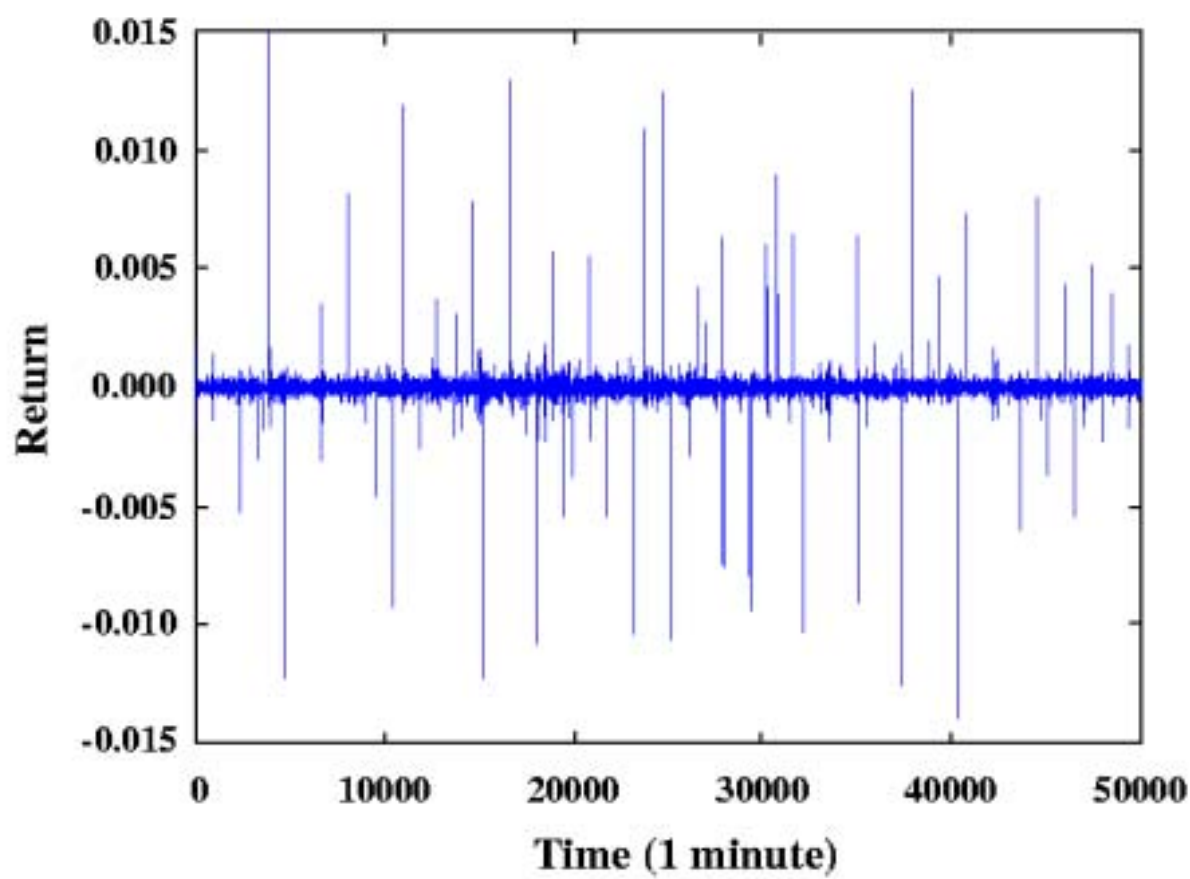

**Fig. 1** Plot of the price return $R(t) = \ln[P(t+\tau)/P(t)]$ for minutely time series of the yen-dollar exchange rate.

To present the averaged distribution of cluster, let us suppose three return states composed by $N$ agents, i.e. the continuous tick data of the yen-dollar exchange rate. We assume that from the states of agent $l$ composed of the three states $\varphi_l = \{-1, 0, 1\}$, where the state of clusters is represented in terms of

$$S(t) = \sum_{i=1}^{N} \varphi_l. \qquad (2)$$

Here the waiting state that occurs no transactions or gets no return corresponds to $\varphi_l = 0$, and the selling and buying states, i.e. the active states of transaction, are $\varphi_l = 1$ and $\varphi_l = -1$, respectively. Assuming that it belongs to the same cluster between a group of agents sharing the same information and making a common decision, the active states of transaction can be represented by vertices in a network having links of time series.

The characteristic feature of herds can be described by the probability distribution of returns $P(R)$. To find the distribution of the price return for different herding probabilities, the herding parameter of the network of agents can be estimated from $P(\varphi_l = +1, -1) = a = a_+ + a_-$, where $P(\varphi_l = +1) = a_+$ and $P(\varphi_l = -1) = a_-$ are, respectively, the probability of the selling and buying herds. In reality we can perform the simulation of $P(R)$ by using the herding parameter $h = (1-a)/a$, i.e. the ratio of no herding probabilities to herding probabilities. The herding parameter is incorporated into the price return, whose elements are usually the random numbers proportional to the real data in Fig. 1. As it analyzes trading tick data on yen-dollar exchange rate, the probability distribution of returns for three herding parameters satisfies the power law

$$P(R) \sim R^{-\beta} \qquad (3)$$

with the scaling exponent $\beta = 3.11$ ($\tau = 1$ minute), 2.81 (30 minutes), and 2.29 (1 hour) in Figs. 2, 3, and 4. The exponents for the won-dollar exchange rate and the KOSPI are, respectively, $\beta = 2.2$ and 2.4 ($\tau = 1$ day) [18,19]. The crash regime appears to increase in the probability of high returns values, since the state of the transaction exists to decrease lesser. In particular we can find the critical herding parameter $h^* = 2.33$ ($a = 0.3$) from our minutely tick data, similar to the case for the won-dollar exchange rate and the KOSPI [19]. From Figs. 2 - 4, it is especially found that it occurs no crash at $\tau > 30$ minutes, and that it occurs the phase transition at $\tau = 30$ minutes.

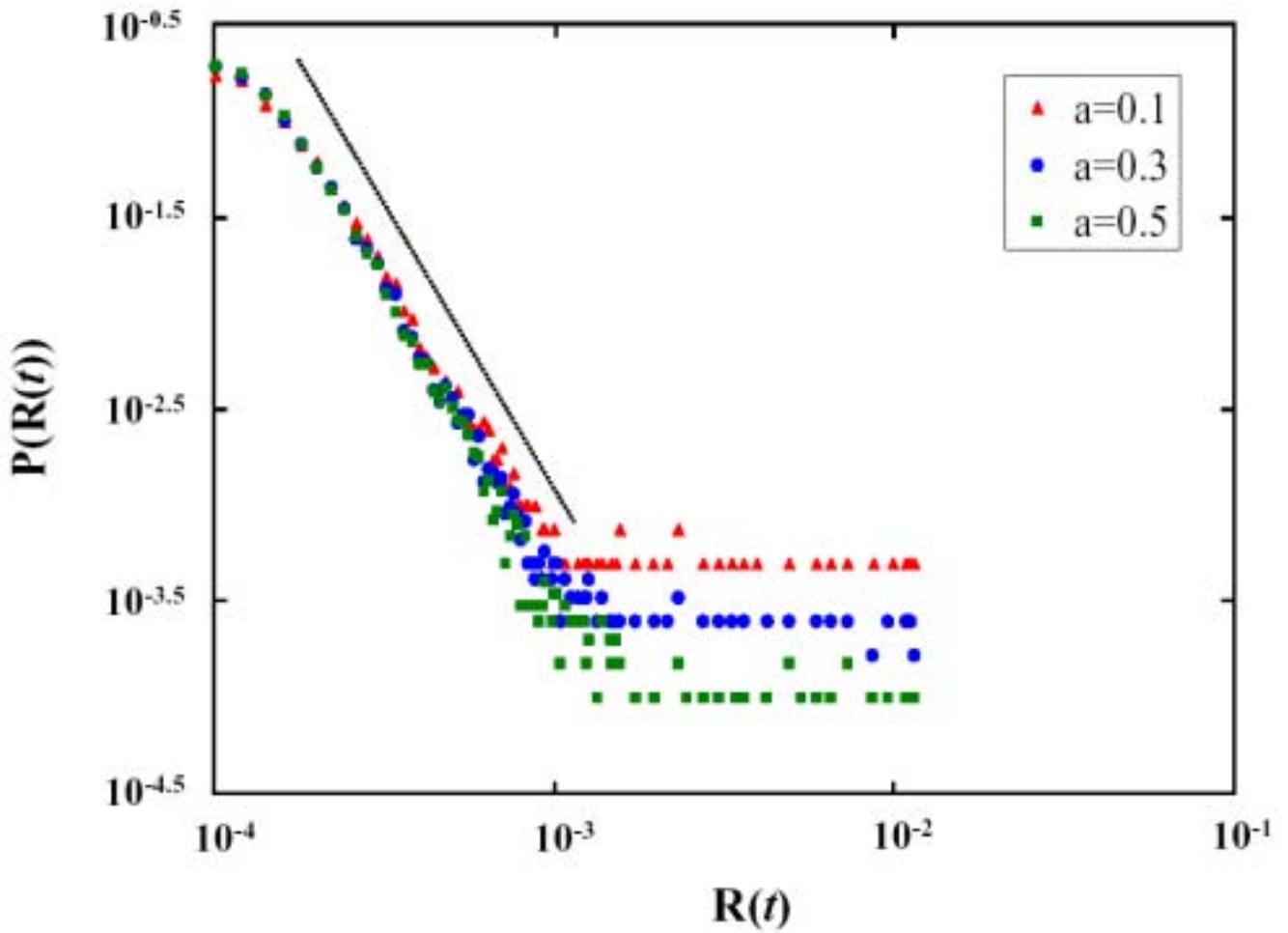

**Fig. 2** Probability distribution of returns for three types of herding probabilities $a = 0.1$, 0.3, 0.5 ($h = 9$, 2.33, 1), where the dotted line scales as a power law $R^{-\beta}$ with the exponent $\beta = 3.11$ ($\tau$ = one minute).

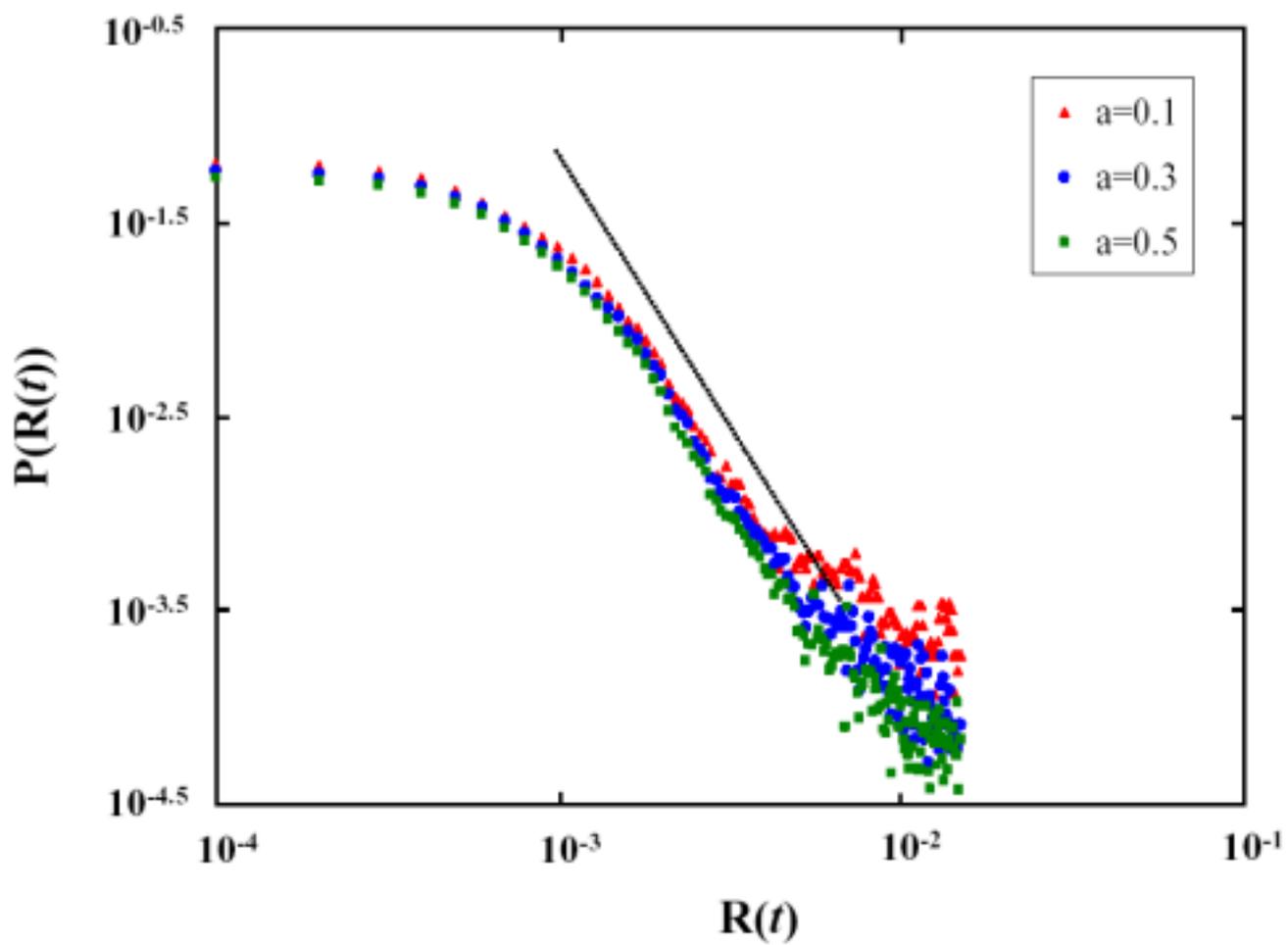

**Fig. 3** Log-log plot of the probability distribution of returns for three types of herding probabilities, where the dotted line scales as a power law with the exponent $\beta = 2.81$ ($\tau = 30$ minutes) and the phase transition is occurred at one time lag $\tau = 30$ minutes.

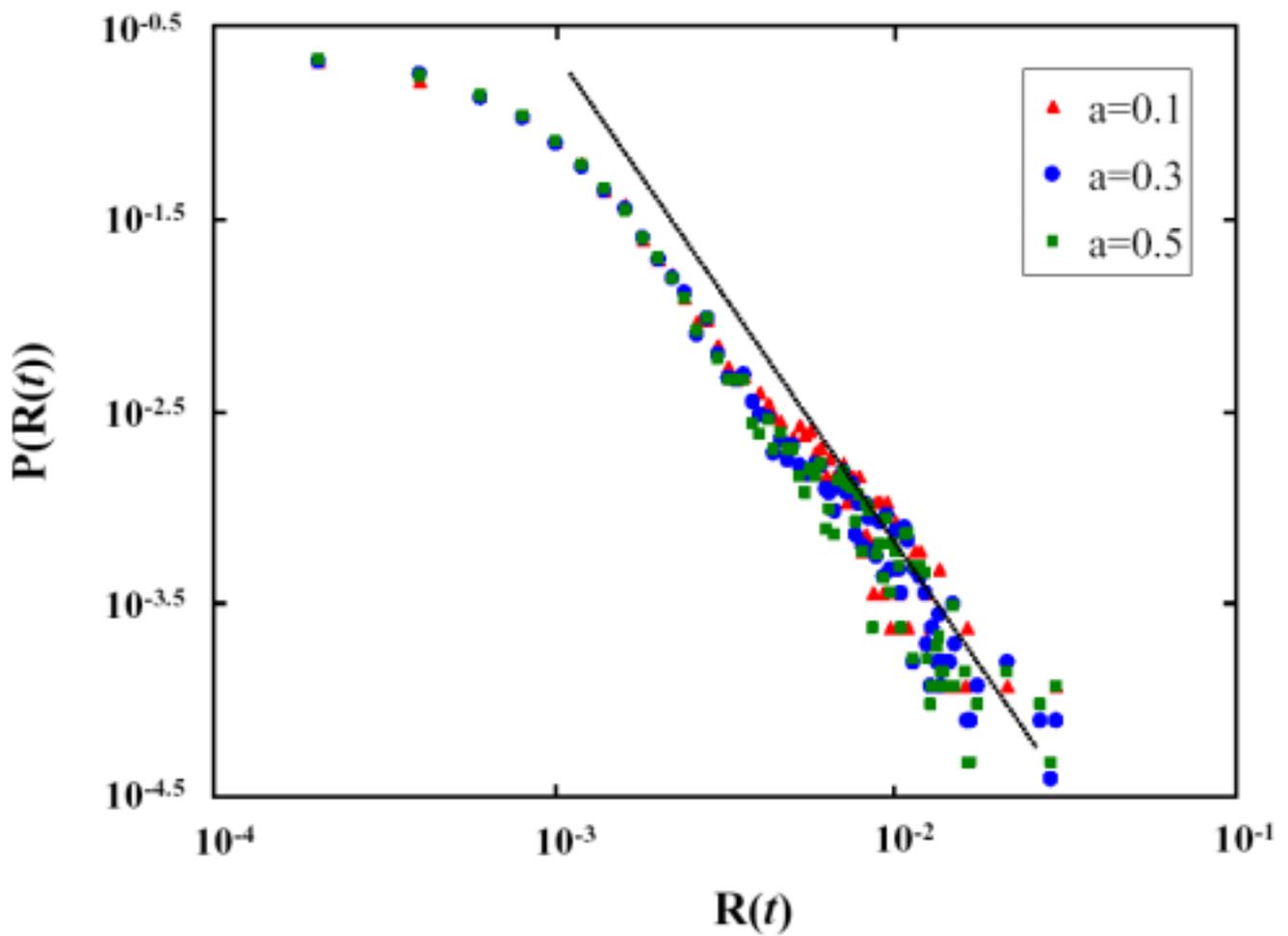

**Fig. 4** Plot of the probability distribution of returns for three types of herding probabilities $a$ =0.1, 0.3, 0.5 ($h$ =9, 2.33, 1), where the dotted line scales as a power law $R^{-\beta}$ with the exponent $\beta = 2.29$ ($\tau = 1$ hour).

In conclusions, we have investigated the dynamical herding behavior for the yen-dollar exchange rate in Japanese financial market. The distribution of the price return scales as a power law $P(R) \sim R^{-\beta}$ with the exponents $\beta$ = 3.11 ($\tau$ = 1 minute), 2.81 (30 minutes), and 2.29 (1 hour), similar to the scaling exponents for the won-dollar exchange rate [19]. However, our distributions of the price return are not in good agreement with the other result [6], but it is in practice found that our scaling exponents are somewhat larger than the numerical 1.5. It would be noted that the probability existing financial crashes is high, because the active herding behavior occurs with the increasing probability as the herding parameter becomes larger value in real financial Markets. We can find the critical herding parameter *h\** = 2.33 from our minutely tick data, similar to the case for the won-dollar exchange rate and the KOSPI. It is obtained that the financial crashes occur at *a* < 0.3 (*h* > 2.33). We would suggest that the phase transition of dynamical herd behaviors is occurred at one time lag $\tau$ = 30 minutes. Our analysis plans to investigate in detail the herd behavior for other foreign exchange rates, and we also hope that the dynamical herd behaviors apply to the tick data of many options in financial markets.

## Acknowledgments

This work was supported by Korea Research Foundation Grant (KRF-2004-002-B00026).